\documentclass[letterpaper,english,aps, nofootinbib, pr, twocolumn, lengthcheck, superscriptaddress]{revtex4-2}

\usepackage[T1]{fontenc}
\usepackage[colorlinks=true,linktocpage=true,linkcolor=blue,citecolor=blue]{hyperref}
\usepackage{amsmath}
\usepackage{color,txfonts}
\usepackage{natbib}
\usepackage{verbatim}
\usepackage{bm}
\usepackage{amssymb}
\usepackage{graphicx}
\usepackage{slashed}
\usepackage{wrapfig}
\usepackage{bbm}
\usepackage[caption=false]{subfig}
\usepackage{verbatim}

\begin{document}

\title{The Non-parametric Equation of State Realizes a Generalized Quark-Hadron Crossover}

\author{Yong-Jia Huang}
\email{huangyj@pmo.ac.cn}
\affiliation{Key Laboratory of Dark Matter and Space Astronomy, Purple Mountain Observatory, Chinese Academy of Science, Nanjing, 210023, China}
\affiliation{RIKEN Center for Interdisciplinary Theoretical and Mathematical Sciences (iTHEMS), RIKEN, Wako 351-0198, Japan}

\author{Shao-Peng Tang}
\affiliation{Key Laboratory of Dark Matter and Space Astronomy, Purple Mountain Observatory, Chinese Academy of Science, Nanjing, 210023, China}

\author{Yi-Zhong Fan}
\affiliation{Key Laboratory of Dark Matter and Space Astronomy, Purple Mountain Observatory, Chinese Academy of Science, Nanjing, 210023, China}
\affiliation{
School of Astronomy and Space Sciences, University of Science and Technology of China, Hefei, Anhui 230026, China.}

\begin{abstract}
We propose a non-parametric approach to construct the statistical equation of state (EOS) continuously from the nuclear crust to the asymptotic-freedom regime. The combined requirement of supporting two-solar-mass neutron stars (NSs) with relatively small radii at low masses and of reaching the asymptotically soft pQCD boundary forces a squared-sound-speed ($c_s^2$) peak followed by extended softening, with $c_s^2$ returning toward $1/3$ near $\sim\!30\,n_{\rm sat}$. Correspondingly, the trace anomaly $\Delta \equiv 1/3 - p/\epsilon$ becomes positive beyond NS densities and approaches the pQCD limit from above. By quantifying the degree of this softening in the posterior, we find evidence for a hadron-quark transition in the cores of the most massive neutron stars. More importantly, this shows that the thermodynamic structure of a complete crust-to-pQCD statistical EOS naturally realizes a generalized quark-hadron crossover. The quark EOS above NS densities is therefore soft and non-perturbative, in contrast to the stiff quark EOS underlying the quark-star picture.
\end{abstract}

\maketitle

{\it Introduction.} Nuclear interactions at densities far above saturation are difficult to probe in terrestrial experiments, but such extreme conditions occur naturally in the Universe in neutron stars. Observations of neutron stars constrain their global structure and, in turn, the thermodynamic properties of cold, dense matter in their interiors, commonly encoded in the equation of state (EOS), i.e., the relation between pressure $p$ and energy density $\epsilon$. In the past few years, multi-messenger measurements---most notably the tidal deformability inferred from the binary-neutron-star merger GW170817 \cite{LIGOScientific:2018cki,LIGOScientific:2018mvr} and mass--radius constraints \cite{Vinciguerra:2023qxq,Salmi:2024aum,Dittmann:2024mbo,Choudhury:2024xbk,Mauviard:2025dmd} from X-ray pulse-profile modeling by the Neutron Star Interior Composition Explorer (NICER)---have substantially improved the accuracy with which neutron-star properties can be determined. These observations are beginning to constrain not only the stiffness of dense matter, but also how matter evolves from hadronic to quark degrees of freedom inside neutron stars.

\begin{figure}[htbp]
\centering
\vspace{-0.3cm}
\includegraphics[width=0.45\textwidth]{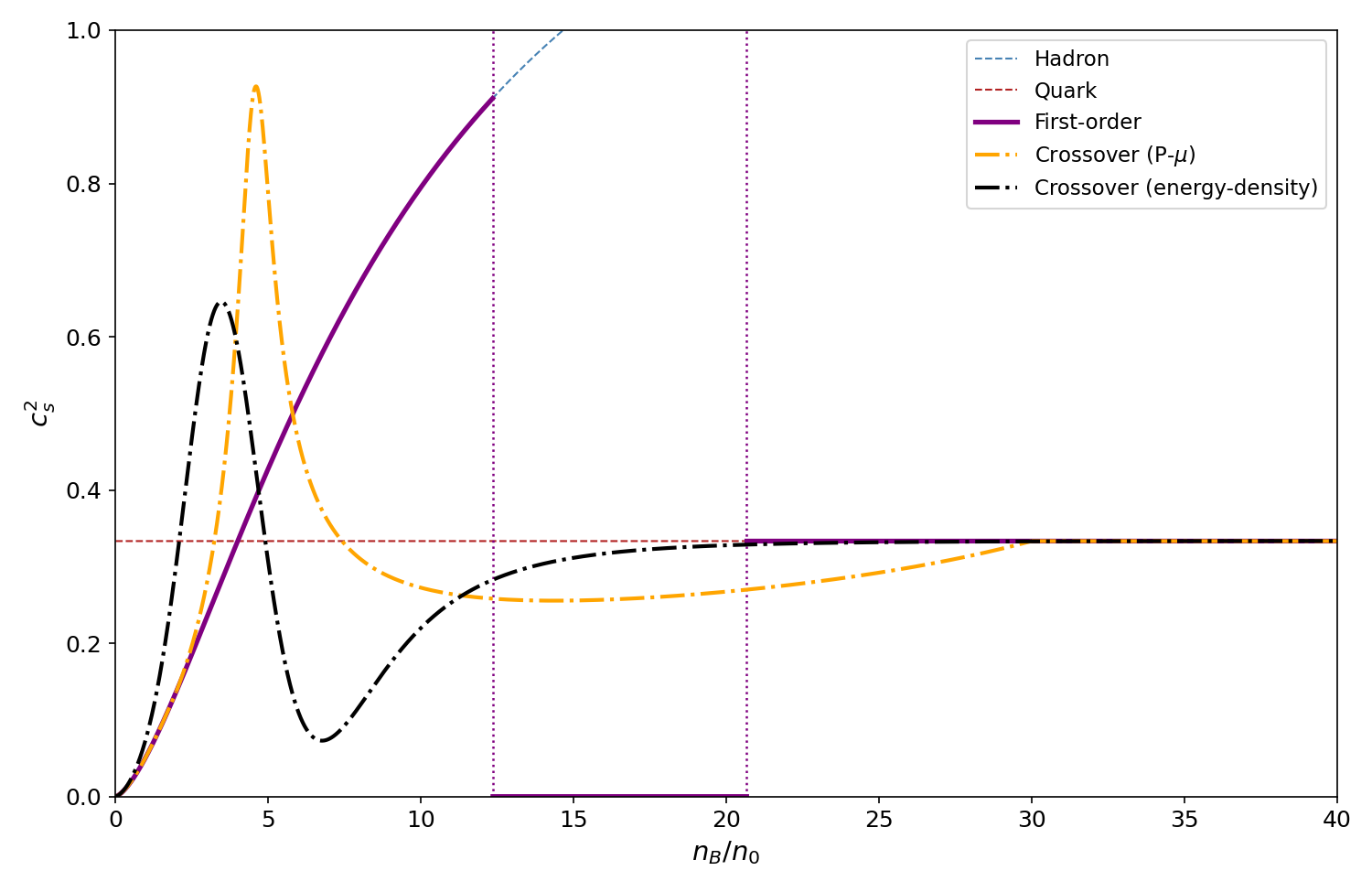}
\vspace{-0.3cm}
\caption{Schematic comparison of hybrid models based on the same hadronic (polynomial) and quark (MIT bag model, $B = 210$\,MeV) EOS. The first-order transition occurs at the chemical potential where $P_H = P_Q$. For the crossover constructions, a fixed-boundary version uses a fifth-order polynomial in $P$--$\mu$ space, matched to pure quark matter at $30\,n_{\rm sat}$ to connect with the asymptotic-freedom regime, while a boundary-free version interpolates the energy density via $w(n) = (n/n_c)^\gamma / [1 + (n/n_c)^\gamma]$ with $n_c = 5\,n_{\rm sat}$ and $\gamma = 4$. The $c_s^2$ peak is introduced to mimic the stiffening required for a soft hadronic EOS to support $2\,M_\odot$ neutron stars.}
\label{fig:schematic_crossover}
\end{figure}

Constructing the hadron-to-quark EOS requires a prescription for how the transition proceeds. The best-known realization is a first-order phase transition via the Maxwell construction, which produces a density discontinuity and a vanishing sound speed at the phase boundary. An alternative is a smooth crossover at finite baryon density, analogous to the one established at finite temperature, constructed either with a fifth-order polynomial in $P$--$\mu_B$ space between fixed hadron and quark densities\cite{Baym:2019iky,Kojo:2021ugu} or by interpolating the energy density of the two phases with a switching function\cite{Masuda:2015kha} (Fig.~\ref{fig:schematic_crossover}). The statistical EOS construction addresses the same physical connection from a different direction. It is not a microscopic hybrid model itself, but a framework for sampling thermodynamically admissible EOSs under given physical constraints. We therefore ask whether the microscopic and statistical constructions share a common mathematical structure, as the special and the general cases of the same hadron-to-quark connection.

At low densities, cold matter is described by nucleons and nuclear interactions with controlled input from chiral effective field theory ($\chi$EFT) \cite{Drischler:2017wtt,Drischler:2020hwi}, while at asymptotically high densities asymptotic freedom makes quarks the appropriate degrees of freedom and imposes a rigorous perturbative-QCD (pQCD) constraint \cite{Gorda:2021znl,Komoltsev:2021jzg}. Massive NS cores probe the intermediate densities where additional degrees of freedom such as hyperons or deconfined quarks may appear \cite{Annala:2019puf,Bauswein:2025dfg}. These unknowns limit the reliability of purely ab initio extrapolations between the two well-controlled regimes. Consequently, microphysical inferences from observations are often drawn within parameterized EOS families or specific phenomenological models \cite{Read:2008iy,Lindblom:2013kra,Annala:2021gom,Grundler:2026zus} and remain conditional on the underlying assumptions. Non-parametric EOS inference \cite{Landry:2018prl,Fujimoto:2019hxv,Landry:2020vaw,Han:2021kjx} was developed precisely to overcome this limitation by providing flexible EOS representations. By combining astrophysical constraints with the low-density $\chi$EFT input and the high-density pQCD information, these approaches bound the set of thermodynamically admissible EOSs \cite{Gorda:2022jvk,Han:2022rug,Annala:2023cwx,Tang:2025xib}.

Since the pQCD regime lies far above typical NS densities, standard non-parametric methods, such as Gaussian Processes (GPs)\cite{Landry:2018prl,Tang:2025xib}, often struggle to connect these widely separated density regimes. A GP with a simple global mean function tends to suppress the sustained structural change needed to connect a stiff intermediate-density EOS to the near-conformal pQCD limit. Recent studies have incorporated high-density information by extrapolating the pQCD constraints down to lower densities under causality and thermodynamic stability \cite{Komoltsev:2021jzg}, but this has led to a debate regarding the density of pQCD likelihood implementation \cite{Somasundaram:2022ztm, Fan:2023spm, Brandes:2023hma}. The inferred properties of the NS core, such as whether the EOS systematically soften, depend sensitively on the density at which the pQCD constraint is imposed \cite{Komoltsev:2023zor}.

In this work, we remove this ambiguity by constructing a non-parametric EOS ensemble that extends continuously from the nuclear crust to the pQCD limit at $\mu_B=2.6\,\mathrm{GeV}$. To overcome the limitations of standard GP priors, we introduce an action-optimized intermediate segment, where the candidate $c_s^2(n)$ paths are assigned an action based on their mismatch from a linear interpolation between the endpoint states, and the path minimizing this action is selected from several physically motivated classes and then projected onto the exact endpoint constraints via the GP covariance. This yields a continuous EOS from the crust to the pQCD anchor and allows the impact of the asymptotic QCD boundary on neutron-star interiors to be determined without introducing an arbitrary implementation density. We then show that a thermodynamically self-consistent statistical EOS shares the same mathematical structure as a generalized quark-hadron crossover. Both of them require a sound-speed peak followed by extended softening to connect the hadronic regime to the asymptotically soft quark regime while supporting massive neutron stars. The quark matter above NS density is therefore soft and non-perturbative.

\begin{figure}[htbp]
\centering
\vspace{-0.3cm}
\includegraphics[width=0.45\textwidth]{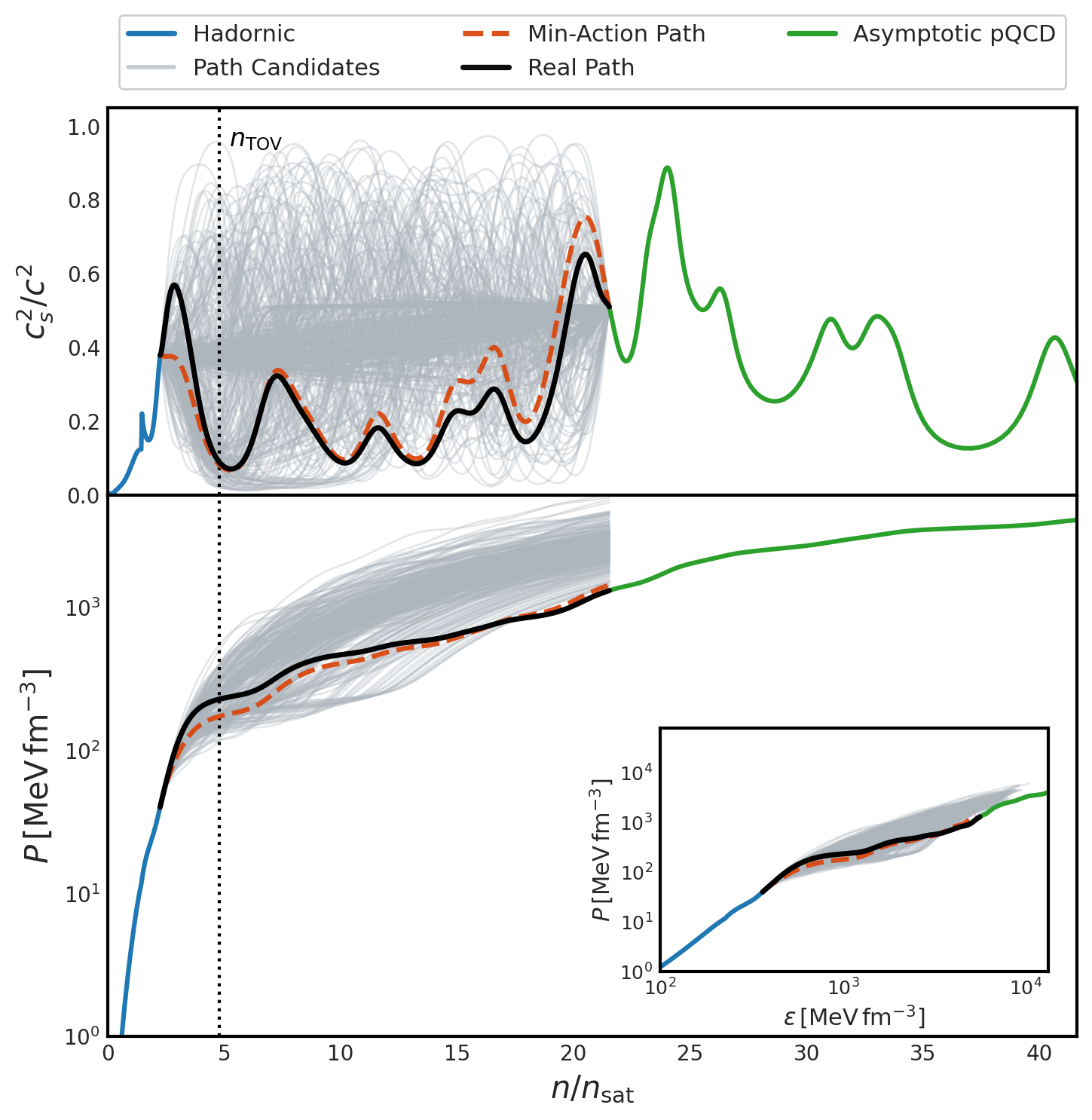}
\vspace{-0.3cm}
\caption{EOS construction in the $c_s^2$--$n$ plane (top) and the $p$--$n$ plane (bottom). The statistical EOS is divided into three parts: (i) {\it hadronic} segment, constrained by low-density nuclear theory and its extension (blue); (ii) {\it Asymptotic pQCD} segment, anchored at $\mu = 2.6\,\mathrm{GeV}$ and extended downward by reverse construction (green); and (iii) {\it Action-Optimized} segment, represented by candidate bridge paths in the intermediate-density region (gray). The selected least-action path, which yields a globally continuous EOS, is shown in dashed red. Black curve is the real path after the condition.}
\label{fig:schematic}
\end{figure}

{\it Methods.} The dense-matter EOS spans multiple orders of magnitude, connecting strongly coupled nuclear matter at low density to weakly coupled quark matter at asymptotically high density. Since ab initio methods are reliable only at these two extremes, constructing a complete cold EOS from the crust to the pQCD regime is a thermodynamic boundary-value problem. This is the same problem faced by the microscopic hybrid constructions discussed in Fig.~\ref{fig:schematic_crossover}. The non-parametric statistical approach addresses it in a model-agnostic way, without prescribing how the two regimes are connected.

The mathematical difficulty is that the thermodynamic integration follows $d\epsilon/dn = (\epsilon + P)/n$, so even small variations in $c_s^2(n)$ at intermediate densities accumulate nonlocally in the integrated energy density. A path that reproduces the endpoint pressure can therefore still miss the required $\epsilon$, since both are coupled through the same $c_s^2(n)$ profile. Forward-shooting methods \cite{Annala:2019puf,Altiparmak:2022bke,Jiang:2022tps} discard paths that miss the target, but are typically built on parametric models whose rigid forms introduce structural biases. Extrapolating the pQCD constraints downward under causality and thermodynamic stability \cite{Komoltsev:2021jzg,Gorda:2022jvk,Han:2022rug} can instead be paired with non-parametric representations, yet it ignores the physics above the matching point and turns the matching density into a free parameter \cite{Somasundaram:2022ztm,Fan:2023spm,Brandes:2023hma}.
A recent Brownian-bridge construction \cite{Gorda:2025aiu} bridges the two endpoints directly, though its diffusion correlation length remains a free prior parameter. In all cases, the method must resolve the same boundary-value problem while avoiding structural bias in the intermediate region.

We represent the EOS through the squared sound speed $c_s^2(n)$ and the Gaussian Process (GP) is placed on the auxiliary field $\phi(n) = -\ln(1/c_s^2(n) - 1)$ \cite{Landry:2018prl}, so that any real GP realization maps to $0<c_s^2<1$. As illustrated in Fig.~\ref{fig:schematic}, each EOS consists of three smoothly connected segments. A low-density hadronic segment embeds the BPS crust\cite{Baym:1971} and $\chi$EFT input\cite{Drischler:2017wtt,Drischler:2020hwi}. A high-density pQCD segment is anchored at $\mu_{\rm pQCD}=2.6\,\mathrm{GeV}$ and integrated downward. The intermediate action-optimized bridge connects the two segments.

The primary technical challenge lies in formulating the bridge segment. A local interpolation of $c_s^2(n)$ is insufficient, because the endpoint $P$ and $\epsilon$ are determined by the nonlocal accumulation along the entire path. The task is thus an inverse boundary-value problem, in which one must find a continuous $c_s^2(n)$ trajectory connecting the hadronic and pQCD endpoint in the coupled $(P,\epsilon)$ space.

Without imposing a fixed transition prescription, we use an action-optimized path selection, motivated by the principle that in many stochastic and variational systems the most probable path between fixed boundary states minimizes an action $S=\int L\,dt$ \cite{FreidlinWentzell2012,Touchette:2009mis}. For each pair of endpoints, we generate candidate $c_s^2(n)$ profiles from broad structural classes (peak-like, first-order-like, monotonic, and GP-template), integrate each through the thermodynamic equations, and evaluate how coherently it approaches the high-density boundary.

The action functional respects the thermodynamic covariance between $P$ and $\epsilon$. We define normalized deviations $\Delta P(n) = [P(n)-P_{\rm tar}(n)]/P_{\rm high}$ and $\Delta\epsilon(n) = [\epsilon(n)-\epsilon_{\rm tar}(n)]/\epsilon_{\rm high}$ relative to a target trajectory connecting the endpoints, and penalize only the component orthogonal to the target thermodynamic direction,
$\mathcal{S}=\int_{n_{\rm low}}^{n_{\rm high}}
w(n)\left[\Delta P(n)-c_{s,\rm tar}^2\Delta\epsilon(n)\right]^2 dn$ ,
where $c_{s,\rm tar}^2 = (P_{\rm high}-P_{\rm low})/(\epsilon_{\rm high}-\epsilon_{\rm low})$. The weight $w(n)$ rises toward the pQCD endpoint. This leaves the intermediate-density EOS structurally flexible, while requiring the accumulated thermodynamic evolution to align with the high-density boundary.

The minimum-action path defines the bridge trajectory. Because it is selected for its thermodynamic structure rather than by direct endpoint shooting, it retains physically relevant features such as rapid stiffening or softening, at the cost of only approximately satisfying the pQCD endpoint. We therefore project the resulting GP realizations onto the exact pQCD values through GP conditioning in $\phi$-space \cite{Rasmussen2006,jidling2017linearly}. This projection distributes the necessary correction according to the GP covariance, preserving smoothness and avoiding a localized deformation of the path. The final EOS exactly satisfies the coupled pQCD constraints in both $P$ and $\epsilon$, while the intermediate structures remain driven by the physical requirements of massive-star support, low-density nuclear input, and the near-conformal quark boundary.

We evaluate the resulting EOS ensemble using astrophysical likelihoods. We include NICER mass--radius constraints for PSR~J0030+0451 \cite{Vinciguerra:2023qxq}, PSR~J0740+6620 \cite{Salmi:2024aum}, PSR~J0437$-$4715 \cite{Choudhury:2024xbk}, and PSR~J0614$-$3329 \cite{Mauviard:2025dmd}, the GW170817 mass--tidal-deformability posterior \cite{LIGOScientific:2018cki}, and the two-solar-mass pulsar requirement.

\begin{figure}[htbp]
\centering
\vspace{-0.1cm}
\includegraphics[width=0.5\textwidth]{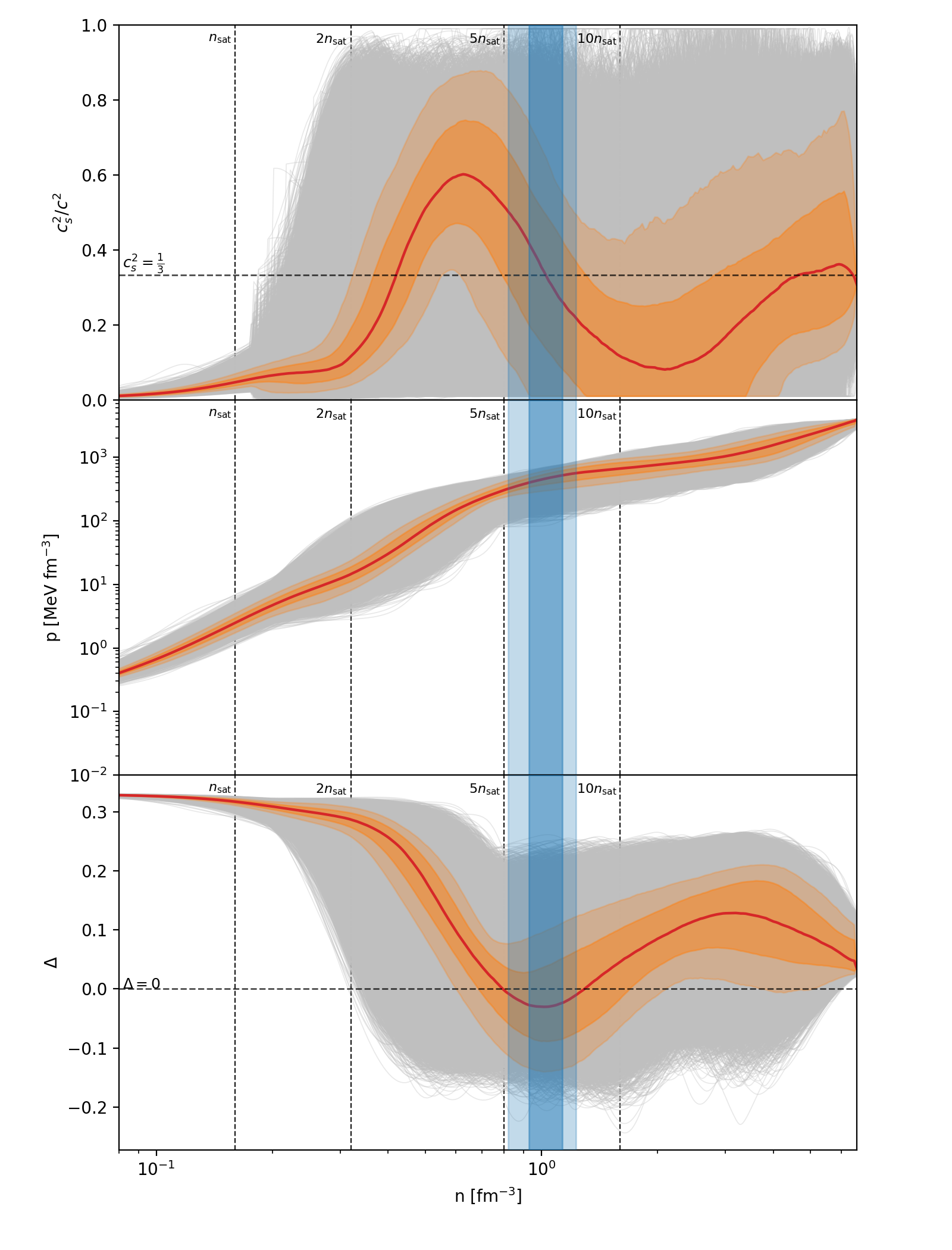}
\vspace{-0.1cm}
\caption{Posterior EOS constraints in the $n$-representation: pressure $p(n)$ (top),
the square of sound speed $c_s^2(n)$ (middle), and normalized trace anomaly
$\Delta(n)=1/3-p/\epsilon$ (bottom), shown with 68\%/95\% credible regions. Gray curves represent the physical priors, and the blue band marks the inferred $n_{\rm TOV}$ range. The joint view highlights a soft--stiff--soft sequence, where the EOS re-softens at high densities to approach the near-conformal pQCD limit ($\Delta \to 0^+$).}
\label{fig:combo-n}
\end{figure}

{\it Results.} The reconstructed posterior EOSs, extending continuously from the low-density $\chi$EFT region to the high-density pQCD limit at $\mu_B = 2.6$~GeV, are shown in Fig.~\ref{fig:combo-n}. The 68\% and 95\% credible intervals for the pressure $p(n)$, the squared sound speed $c_s^2(n)$, and the normalized trace anomaly $\Delta(n)\equiv 1/3-p/\epsilon$ \cite{Fujimoto:2022ohj} are plotted as functions of the baryon density $n$. The blue band marks the posterior range of the central baryon density of the maximum-mass configuration, $n_{\rm TOV}$. 
The posterior favors a non-monotonic sound-speed profile, where $c_s^2$ rises above the conformal value $1/3$ at intermediate densities, as has been studied in quark-hadron crossover models \cite{Masuda:2012kf,Baym:2019iky,Kojo:2021ugu} and in sign-problem-free lattice simulations in QCD-like theories \cite{Iida:2022hyy,Iida:2024irv,Brandt:2022hwy}. The posterior further shows that this peak is followed by substantial softening, with $c_s^2$ returning toward $1/3$ near $\sim\!30\,n_{\rm sat}$ (see also \cite{tang:2026c} for confirmation with an independent approach). Both the peak and the subsequent drop occur mainly below $n_{\rm TOV}$, indicating that the full peak-to-softening structure is realized inside the cores of the most massive NSs .

This result should be contrasted with constructions that impose pQCD through an adjustable matching density \cite{Han:2022rug,Gorda:2022jvk}, as discussed in the Introduction. In those approaches, the required softening can be postponed to densities above $n_{\rm TOV}$, leaving the EOS in the centers of massive stars comparatively stiff. In our framework, each posterior sample is a single thermodynamic trajectory extending to $\mu_B=2.6$~GeV, and the re-softening appears as part of the global crust-to-pQCD connection. The non-monotonic structure of $c_s^2$ therefore truly manifests within the cores of the most massive NSs.

The physical origin of this non-monotonic behavior is the coupled boundary-value constraint. At low density, the trace anomaly is initially positive because hadrons are massive and nuclear interactions are attractive. As density increases, astrophysical observations ($M \ge 2\,M_\odot$) demand a prominent early stiffening of the EOS to prevent gravitational collapse, inherently forcing $\Delta$ to become negative ($c_s^2 > 1/3$). However, the thermodynamic relation $\partial_n\epsilon=(\epsilon+p)/n$ implies that this stiffening rapidly accumulates energy density. To reach the exact pQCD endpoint $(p_{\rm pQCD},\epsilon_{\rm pQCD})$ without overshooting the allowed energy density, the early sound-speed peak must be compensated by extended softening at higher densities.

This thermodynamically required softening naturally drives the trace anomaly back to positive values ($\Delta > 0$) at higher densities, before the EOS approaches the conformal limit. Reaching the positive $\Delta$ requires a positive derivative $\partial\Delta/\partial\epsilon > 0$. In the thermodynamic identity $c_s^2 = 1/3 - \Delta - \epsilon\,(\partial\Delta/\partial\epsilon)$, this slope directly lowers $c_s^2$, leading to the strong softening seen inside the core of a massive NS. When $\Delta$ subsequently falls toward zero from above ($\partial\Delta/\partial\epsilon < 0$), this negative slope appears with a minus sign, thereby increasing $c_s^2$ and possibly creating a second structural peak before the EOS finally settles to $1/3$. Importantly, this multi-stage behavior is a global thermodynamic requirement rather than a feature imposed by any specific microscopic model.

The separation between softening and conformality is essential. As shown by the trace anomaly $\Delta(n)$ in Fig.~\ref{fig:combo-n}, the conformal limit $\Delta\to 0^+$ is approached only around $\sim\!30\,n_{\rm sat}$, far above the density range directly probed by stable NS cores. The soft EOS above NS densities should therefore not be interpreted as an already weakly interacting conformal quark gas. Rather, it is a non-perturbative high-density branch required to connect massive-star support to the asymptotically soft pQCD boundary, and its emergence at densities well below the conformal limit is a thermodynamic signature of a first-order phase transition \cite{Constantinou:2025wxj,Gao:2024lzu} or crossover \cite{McLerran:2018hbz,Kojo:2021ugu} to non-conformal matter. In this sense, the statistical EOS recovers the same thermodynamic structure as a generalized boundary-free quark-hadron crossover, where a sound-speed peak is followed by extended softening as the EOS evolves from the hadronic regime toward the quark regime, without imposing a microscopic crossover prescription.

\begin{figure}[htbp]
\centering
\vspace{-0.3cm}
\hspace*{-3mm}\includegraphics[width=0.45\textwidth]{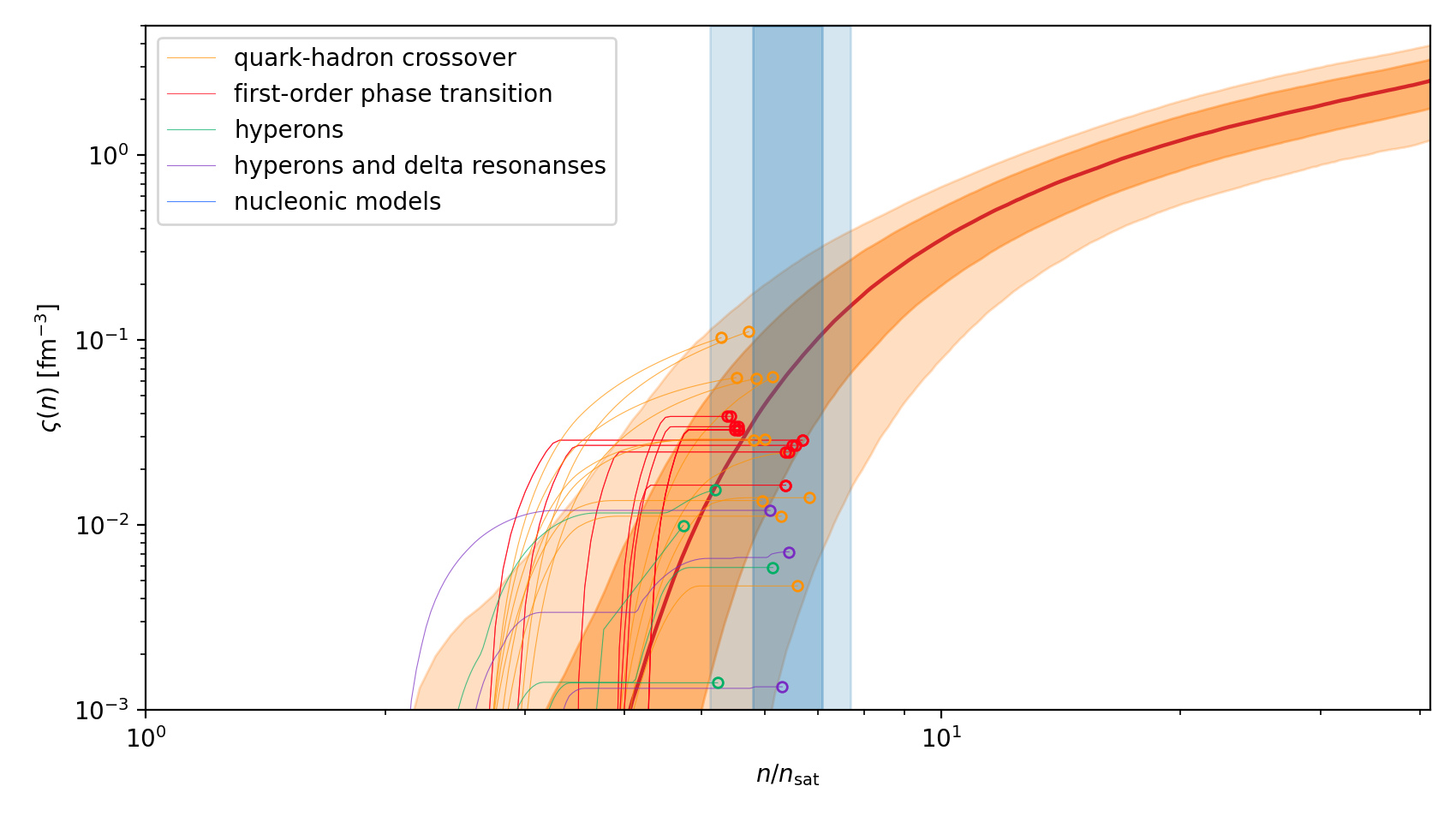} \\
\includegraphics[width=0.45\textwidth]{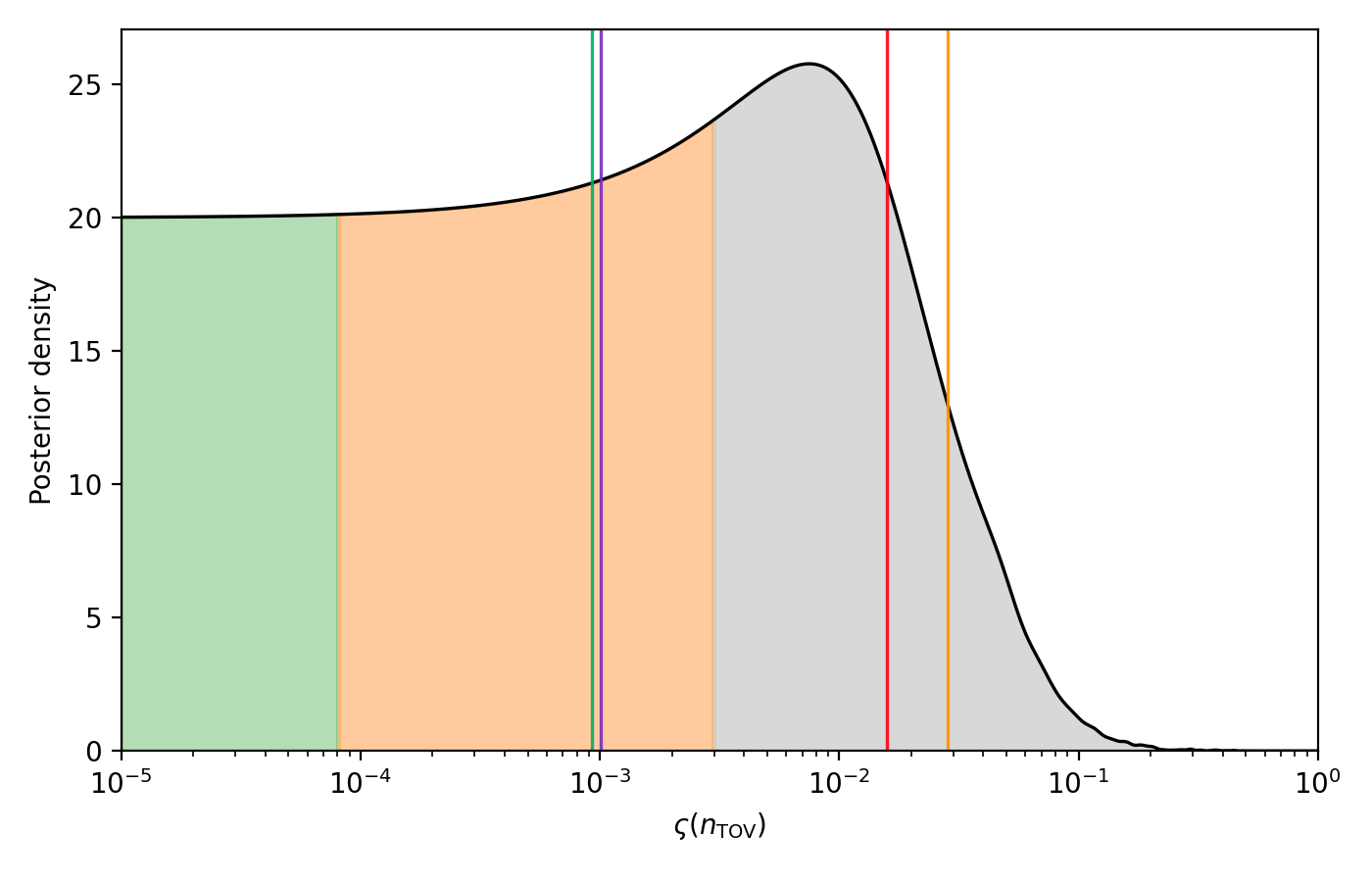}
\vspace{-0.1cm}
\caption{The cumulative softening parameter $\varsigma$ and its posterior distribution. 
Top: $\varsigma$ as a function of the baryon number density $n/n_{\mathrm{sat}}$. The maximum $\varsigma$ for hadronic models from the \texttt{CompOSE} database shown, with another two hadron-quark first-order phase transition and crossover models for comparison. Open circles indicating the termination of each curve at its respective $n_{\rm TOV}$. 
Bottom: Probability density of the cumulative softening index at the maximum density of the star, $\varsigma(n_{\rm TOV})$. The shaded regions show the partition of the posterior confidence intervals (green: below the posterior 95\% lower bound; orange: between the 95\% and 68\% lower bounds; gray: above the 68\% lower bound). Vertical lines represent the values of $\varsigma(n_{\rm TOV})$ for the same models as above. 
Overall, the monotonically increasing trend of the posterior $\varsigma(n)$ within the relevant interior densities underscores a substantial reduction in the $c_s^2$ after reaching its peak. This pervasive softening behavior show the evidence for a hadron-quark transition in the inner cores of massive NSs.}
\label{fig:S_combined}
\end{figure}

To quantify the transition-like character of this softening, we introduce the accumulated softening index $\varsigma$. We define $\mathcal{C}_{s,\rm max}^2(n)=\sup_{n'\le n}\{c_s^2(n')\}$ and measure the accumulated post-peak reduction of the sound speed,  $\varsigma = \int_{n_{\rm min}}^{n_{\rm max}} \left[\mathcal{C}_{s,\rm max}^2(n)-c_s^2(n)\right]^2 dn$.  Fig.~\ref{fig:S_combined} shows the growth of $\varsigma$ with density and its posterior distribution at $n_{\rm TOV}$. For comparison, we evaluate the same quantity for beta-equilibrium EOSs from the \texttt{CompOSE} database \cite{compose}. Even hadronic models with exotic degrees of freedom, such as hyperons and $\Delta$ resonances, produce only limited post-peak softening, with a maximum value near $\varsigma\simeq1.01\times10^{-3}$. The posterior preference for larger accumulated softening therefore supports a hadron-quark transition interpretation in the cores of the most massive NSs. 

We further examine the relation between the intermediate-density sound-speed peak and the maximum mass. Although the posterior favors a pronounced peak, $c_{s,\rm peak}^2=0.68^{+0.14}_{-0.13}$, it shows only a weak correlation with the maximum mass, $M_{\rm TOV}=2.11^{+0.12}_{-0.09}\,M_\odot$ (See Fig.~\ref{fig:mtov_cs2_peak_joint} in Appendix for the details), with a 95\% upper limit of $2.37\,M_\odot$. This decoupling indicates that a pronounced intermediate-density stiffening does not require an exceptionally large $M_{\rm TOV}$ \cite{Shao:2019ioq,Han:2022rug,Fan:2023spm}. 

{\it Summary and discussion.}
In this work, we establish a non-parametric framework that continuously connects the low-density nuclear EOS to the asymptotic pQCD limit, thereby resolving the ambiguity associated with pQCD implementation. Multi-messenger constraints naturally select a non-monotonic sound-speed profile, in which $c_s^2$ peaks well above the conformal value and then undergoes extended softening before reaching the pQCD regime. We introduced an integrated softening index, $\varsigma$, to quantify this post-peak reduction. Compared with purely hadronic models, the inferred softening provides evidence for a hadron-quark transition inside massive neutron stars.

This peak-then-soften structure is not imposed but emerges as a thermodynamic necessity. Supporting massive NSs requires intermediate-density stiffening, yet the same stiffening rapidly accumulates energy density through $\partial_n\epsilon=(\epsilon+p)/n$. To reach the coupled pQCD endpoint without overshooting the allowed energy density, the EOS must soften at higher densities. The statistical EOS therefore naturally realizes the thermodynamic structure of a generalized quark-hadron crossover. Although no strict criterion yet exists to distinguish a first-order phase transition from a smooth crossover using cold EOS information alone, a boundary-free quark-hadron crossover is the most natural interpretation at the level of thermodynamic structure, since it shares the same mathematical structure as the statistical EOS construction. Since the conformal limit $c_s^2 = 1/3$ (as well as $\Delta\to0^+$) is seen reached only after $\sim30\,n_{\rm sat}$, the soft high-density EOS should not be interpreted as a perturbative conformal quark gas, but rather as a soft, non-perturbative quark branch. The quark matter above NS densities is therefore soft, in contrast to the conventional quark-star picture \cite{Witten:1984rs}, which relies on an intrinsically stiff quark EOS to support massive compact stars.

Definitive evidence for the nature of this transition requires probing the finite-temperature properties of dense matter, which can help reveal the underlying QCD phase structure. The extended softening identified here may arise from several different microphysical mechanisms, such as pressure dilution in an extended quark-hadron mixed phase \cite{1992PhRvD..46.1274G}, non-conformal pQCD corrections to quark-gluon interactions \cite{1977PhRvD..16.1169F}, or symmetry-breaking effects from finite strange-quark mass \cite{Fraga:2004gz,Andersen:2026sjc}. Disentangling these possibilities is difficult with only cold, isolated neutron stars. However, cold EOSs that look identical at zero temperature can arise from different microscopic mechanisms and exhibit distinct finite-temperature behavior. Numerical simulations of binary neutron-star mergers \cite{Qiu:2025ybw,Haque:2022dsc,Blacker:2023afl,Huang:2022mqp,Fujimoto:2022xhv,Most:2022wgo,Harada:2023eyg,Hensh:2024onv,Miao:2024qik} provide a critical opportunity to distinguish them. The extreme thermodynamic conditions reached during these highly dynamical events may expose distinct multi-messenger signatures, ultimately allowing us to identify the microphysics driving the transition to deconfined quark matter.

%%%%%%%%%%%%%%%%%%%%%%%%%%%%%%%%%%%%%%%%%%%%%%%%%%%%%%%%%%%
\begin{acknowledgments}
The large-sample EOS construction were carried out on the Hokusai Bigwaterfall supercomputer in RIKEN. This work is supported by the National Natural Science Foundation of China under Grants (No. 12588101, No. 12233011 and No. 12303056), the Project for Young Scientists in Basic Research (No. YSBR-088) of the Chinese Academy of Sciences,
and the Postdoctoral Fellowship Program (No. GZC20241915) of the China Postdoctoral Science Foundation. 
\end{acknowledgments}
%%%%%%%%%%%%%%%%%%%%%%%%%%%%%%%%%%%%%%%%%%%%%%%%%%%%%%%%%%%

\appendix   

\section{EOS Construction}
\label{sec:sm_gp_path}

\subsection{GP representation}
Following the non-parametric strategy introduced in many works \citep{Landry:2018prl,Essick:2019ldf,Landry:2020vaw,Han:2022rug,Tang:2025xib}, we model the EOS through the auxiliary variable
\begin{equation}
\phi(n) \equiv -\ln\!\left(\frac{1}{c_s^2(n)} - 1\right) = \mathrm{logit}[c_s^2(n)],
\end{equation}
which maps the causal range $c_s^2 \in (0,1)$ onto $\phi \in (-\infty, +\infty)$, making it a natural target for GP modeling. The inverse relation recovers the $c_s^2$ via the sigmoid function
\begin{equation}
c_s^2(n) = \frac{1}{1+e^{-\phi(n)}}.
\end{equation}

Before any conditioning, a GP prior is placed on $\phi$ as a function of baryon number density $n$. To construct the mean function of the GP, we establish a density-dependent skeleton $c_{s,\rm skel}^2(n)$. This skeleton acts as a baseline template, interpolating the structural expectations of the EOS—such as tracking the center of the $\chi$EFT band at low densities and transitioning to predefined plateaus (e.g., $c_s^2 =  1/3$) at higher densities. This skeleton is then mapped to the auxiliary space to define the prior mean:
\begin{equation}
\bar\phi(n) = \mathrm{logit}[c_{s,\rm skel}^2(n)].
\end{equation}

The covariance is governed by the Gaussian kernel $K(n,n') = \eta\exp\!\left[-(n-n')^2/(2\ell^2)\right]$. The hyperparameters---the variance $\eta$ and the correlation length $\ell$---are uniformly drawn per EOS sample from their respective hierarchical priors: $\ell \sim \mathcal{N}\!\left(1.0\,n_{\rm sat},\,(0.25\,n_{\rm sat})^2\right)$ and $\eta \sim \mathcal{N}(1.25,\;0.2^2)$, where $n_{\rm sat} \approx 0.16\;\mathrm{fm}^{-3}$ is the nuclear saturation density.

From a sampled $\phi(n)$, we reconstruct the full thermodynamic state. Using the first law of thermodynamics at zero temperature, the baryon chemical potential is $\mu(n) = [\epsilon(n)+p(n)]/n$, and the speed-of-sound definition $c_s^2 = dp/d\epsilon$ yields the coupled system of first-order ODEs:
\begin{equation}
\frac{dp}{dn} = c_s^2(n)\,\frac{\epsilon + p}{n} \equiv c_s^2(n)\,\mu(n), \label{eq:dpdn}
\end{equation}
\begin{equation}
\frac{d\epsilon}{dn} = \frac{\epsilon + p}{n} \equiv \mu(n). \label{eq:dedn}
\end{equation}
These equations are integrated from specified initial conditions $(\epsilon_0, p_0)$ at the lower density boundary using a Heun scheme, ensuring second-order accuracy.

\subsection{Three-domain composite construction}
The full EOS consists of three structurally independent but sequentially continuous segments: the {\it Hadronic} segment, the {\it pQCD} segment, and the action-optimized segment.

\subsubsection{{\it Hadronic} Segment}
\label{sec:domain_low}

Below $\sim 0.3 n_{\rm sat}$, the EOS is imported from the BPS crust table\cite{Baym:1971}. From $0.3 n_{\rm sat}$ onwards, we construct a single GP realization covering the entire range up to the end of the segment (denoted $n_{\rm path,start}$, which serves as the left boundary of the optimization bridge).

The EOS from $0.3 n_{\rm sat}$ to $n_{\rm path,start}$ is divided into two physically distinct zones sharing a single covariance matrix:
\begin{itemize}
\item \textbf{Zone~A} ($0.3 n_{\rm sat} \le n \le n_{\rm CEFT}$): the $\chi$EFT-constrained region, with $n_{\rm CEFT}\sim 1.1\!-\!2.0\,n_{\rm sat}$ sampled uniformly per construction;
\item \textbf{Zone~B} ($n_{\rm CEFT} < n \le n_{\rm path,start}$): the standard GP extrapolation region, where $n_{\rm path,start}$ is sampled from a log-Beta distribution over a configurable range ($2 \, n_{\rm sat}$ to $10 \, n_{\rm sat}$) for efficiency.
\end{itemize}
Crucially, both zones are drawn from the same multivariate normal distribution in a single construction, ensuring analytical continuity and smoothness across the $n_{\rm CEFT}$ boundary. 

The deterministic skeleton in Zone~A is centered on the midpoint of the N3LO $\chi$EFT band \cite{Drischler:2017wtt}: $c_{s,\rm skel}^2(n) = \tfrac{1}{2}\!\left[c_{s,\rm q025}^2(n) + c_{s,\rm q975}^2(n)\right]$, where $c_{s,\rm q025}^2$ and $c_{s,\rm q975}^2$ are the lower and upper 95\% credible limits of the $\chi$EFT calculation interpolation. In Zone~B, the skeleton transitions smoothly (via a cubic Hermite function $\mathcal{H}(t) = t^2(3-2t)$) from the Zone~A endpoint value to a configurable plateau value of 0.5. 

Instead of treating the theoretical band as scattered observation points via standard GP regression, we directly reshape the \emph{prior} covariance to embed the target bandwidth. The 95\% CI for the $\chi$EFT calculation is converted to $\phi$-space:
\begin{equation}
\sigma_{\phi,{\rm target}}(n) = \frac{\mathrm{logit}(c_{s,\rm q975}^2) - \mathrm{logit}(c_{s,\rm q025}^2)}{2\,z_{0.975}},
\label{eq:sigma_target}
\end{equation}
where $ z_{0.975} \approx 1.95996$. A density-dependent scaling array $s(n) = \sigma_{\phi,\rm target}(n) / \sigma_\phi$ is applied inside the $\chi$EFT coverage, and $s(n)=1$ outside. The effective covariance is:
\begin{equation}
K_{\rm eff}(n_i, n_j) = s(n_i)\,K(n_i,n_j)\,s(n_j).
\label{eq:Keff}
\end{equation}
This outer-product construction shapes the prior covariance to natively reproduce the dispersion of the $\chi$EFT uncertainty band, avoiding the matrix inversion instabilities associated with dense pseudo-observations.

The first grid point is pinned exactly to the crust sound speed via zero-noise single-point conditioning:
\begin{equation}
\bm{\mu}_{\rm post} = \bm{\mu}_{\rm prior} + \frac{\bm{k}_0\,(\phi_0^{\rm anchor} - \bar\phi_0)}{K_{00}}, \quad \bm{K}_{\rm post} = \bm{K}_{\rm eff} - \frac{\bm{k}_0\,\bm{k}_0^\top}{K_{00}},
\end{equation}
where $\bm{k}_0 = K_{\rm eff}(n, 0.3 n_{\rm sat})$ and $K_{00} = K_{\rm eff}(0.3 n_{\rm sat},0.3 n_{\rm sat})$. The entire \emph{Hadronic} segment is drawn as $\bm{\phi}_{\rm low} \sim \mathcal{N}(\bm{\mu}_{\rm post},\;\bm{K}_{\rm post})$.

\subsubsection{{\it pQCD} Segment}
\label{sec:domain_pqcd}

At the high-density limit, the EOS is anchored by pQCD evaluated at a baryon chemical potential of $\mu_{\rm pQCD} = 2.6\;\mathrm{GeV}$. To account for the truncation uncertainty, the physical renormalization scale is varied as $\Lambda = X \mu$. For each EOS sample, the dimensionless scale parameter $X$ is drawn from a uniform linear prior $X \sim U(0.5, 2.0)$. 

The EOS variables $(n, p, \epsilon, c_s^2)_{\rm pQCD}$ are computed using state-of-the-art N$^3$LO results for cold quark matter \cite{Gorda:2023mkk}. The pressure of $N_f=3$ massless quark matter is expressed as a series in the strong coupling $\alpha_s$:
\begin{equation}
    p_{\rm pQCD}(\mu, X) = p_{\rm FD}(\mu) \left[ 1 + \sum_{i=1}^3 P_i(\alpha_s, X) \right],
\end{equation}
where $p_{\rm FD}(\mu) = \frac{\mu^4}{108\pi^2}$ is the free Fermi-gas pressure. The perturbative coefficients up to three-loop order are explicitly given by:
\begin{align}
    P_1 &= -2\bar{\alpha}_s, \\
    P_2 &= -3\bar{\alpha}_s^2 \left[ \ln(3\bar{\alpha}_s) + 3\ln X + C_2 \right], \\
    P_3 &= 9\bar{\alpha}_s^3 \left[ \frac{11}{12}\ln^2(3\bar{\alpha}_s) - (6.5968 + 3\ln X)\ln(3\bar{\alpha}_s) + c_{30}(X) \right],
\end{align}
where $\bar{\alpha}_s \equiv \alpha_s/ \pi$ and $C_2 \approx 5.0021$. At full N$^3$LO, the scale-dependent coefficient $c_{30}(X)$ reads
\begin{equation}
    c_{30}(X) = 5.1342 + \frac{2}{3}c_0 - 18.284\ln X - \frac{9}{2}\ln^2 X,
\end{equation}
with the non-perturbative constant adopted as $c_0 = -23$ \cite{Gorda:2023mkk}. The coupling $\alpha_s(\mu, X)$ is evaluated at the two-loop level using standard renormalization-group running:
\begin{equation}
    \alpha_s(\mu,X) = \frac{4\pi}{\beta_0 L} \left( 1 - \frac{\beta_1}{\beta_0^2}\frac{\ln L}{L} \right), \quad L = \ln\!\left( \frac{\mu^2 X^2}{\Lambda_{\overline{\mathrm{MS}}}^2} \right),
\end{equation}
where $\beta_0 = 9$ and $\beta_1 = 64$ for $N_f=3$. We adopt the modified reference scale parameter $\Lambda_{\overline{\mathrm{MS}}} = 0.341 \times 1.5 = 0.5115\;\mathrm{GeV}$.

The EOS in the \emph{pQCD} segment is generated by a separate GP operating in the chemical potential $\mu$-domain, extending from $\mu_{\rm pQCD}$ down to $\mu_{\rm path,end}$ (the chemical potential corresponding to $n_{\rm path,end}$). Similar to the {\it Hadronic} segment, we explicitly construct a prior skeleton $c_{s,\rm skel}^2(\mu)$ which smoothly ramps from the conformal value $c_s^2 = 1/3$ at the pQCD boundary to a configurable value $\bar c_{s,\rm end}^2$ at the lower boundary along the normalized coordinate $(\mu - \mu_{\rm cross,end})/(\mu_{\rm pQCD} - \mu_{\rm cross,end})$. We consider scenarios anchoring $\bar c_{s,\rm end}^2$ as $1/3$; both considerations yield consistent inferred results. Because the numerical domain in $\mu$-space (spanning $>1000$ MeV) is vastly larger than the natural width in $n$-space, the GP length scale is dilated to $\ell_\mu = 500\,\ell_n$, ensuring the generated variations maintain corresponding physical smoothness across coordinate parameterizations and preventing unphysical high-frequency numerical oscillations.

The \emph{pQCD} trajectory $\phi(\mu)$ is conditioned at $\mu_{\rm pQCD}$ with $\phi = \mathrm{logit}(c_{s,\rm pQCD}^2)$ via zero-noise point conditioning. The thermodynamic state is reconstructed by \emph{reverse} analytical integration of the coupled ODEs stepping downward to $\mu_{\rm path,end}$. Reversing the arrays yields $(n, \epsilon, p, c_s^2)$ ascending toward the $n_{\rm pQCD}$.

\subsubsection{Action-optimized Segment}
\label{sec:domain_bridge}

The intermediate regime includes the main technical challenge: \emph{both} the starting state $(\epsilon_{\rm path,start}, p_{\rm path,start}, c_{s,\rm path,start}^2)$ and the ending state $(\epsilon_{\rm path,end}, p_{\rm path,end}, c_{s,\rm path,end}^2)$ are strictly connected to the bounding segments. An unconstrained GP sample broadly fails to close this integral gap. We therefore compute the thermodynamic integral over candidates to find the most natural topological path.

To capture the rich variety of potential physics, we generate at least 100 randomly parameterized candidate $c_s^2(n)$ mean profiles for each of four structural families:

\begin{itemize}
\item \textbf{Peak}: a skewed Gaussian structure modeling scenarios with a pronounced sound-speed maximum. The peak is parameterized by its relative center position within the segment ($[0.12,0.88]$), width, kurtosis, and skewness, coupled with a sigmoid-modulated tail to ensure smooth boundary connection;
\item \textbf{First-order (Maxwell/Gibbs)}: a structure mimicking first-order phase transitions. It is constructed as a rapid drop to a local minimum floor, with flat flooring (Maxwell-like) and sloped flooring (Gibbs-like);
\item \textbf{Monotonic}: a continuously stiffening or softening structure. The profile is shaped by a power-law (with scaling exponents drawn uniformly $\in [0.6, 1.8]$) to uniformly explore monotonically increasing or decreasing path;
\item \textbf{GP template}: a direct GP sample explicitly conditioned on the boundary values.
\end{itemize}

For each candidate $c_{s,i}^2(n)$, the corresponding pressure and energy density is integrated via Eqs.~(\ref{eq:dpdn})--(\ref{eq:dedn}) originating from the lower bound. To measure the macroscopic mismatch along the path, a target theoretical baseline $(\epsilon_{\rm tar}(n), p_{\rm tar}(n))$ is defined via direct linear interpolation between the boundary:
\begin{align}
\epsilon_{\rm tar}(n) &= \epsilon_{\rm path,start} + (\epsilon_{\rm path,end} - \epsilon_{\rm path,start}) \frac{n - n_{\rm path,start}}{n_{\rm path,end} - n_{\rm path,start}}, \\
p_{\rm tar}(n) &= p_{\rm path,start} + (p_{\rm path,end} - p_{\rm path,start}) \frac{n - n_{\rm path,start}}{n_{\rm path,end} - n_{\rm path,start}}.
\end{align}

The normalized tracking deviations are defined relative to the absolute scale of the high-density endpoint, 
\begin{equation}
\Delta P(n) = \frac{P(n) - P_{\rm tar}(n)}{P_{\rm high}}, \quad \Delta \epsilon(n) = \frac{\epsilon(n) - \epsilon_{\rm tar}(n)}{\epsilon_{\rm high}}.
\end{equation}
To strictly project out their thermodynamic correlation during tracking, we utilize the target sound-speed squared, evaluated simply by the gradient of the reference paths:
\begin{equation}
c_{s,\rm tar}^2(n) = \frac{\partial_n P_{\rm tar}(n)}{\partial_n \epsilon_{\rm tar}(n)}.
\end{equation}

The thermodynamic action for the profile is evaluated as:
\begin{align}
\mathcal{R}(n) &= \bigl[\Delta p(n) - c_{s,\rm tar}^2(n)\,\Delta\epsilon(n)\bigr]^2, \\
\mathcal{S}_i &= \int_{n_{05}}^{n_{\rm cross,end}} w(n)\,\mathcal{R}(n)\,dn,
\label{eq:action}
\end{align}
where $w(n)$ is the weighting function, modeled as a Half-Gaussian peaked exactly at the high-density boundary:
\begin{equation}
w(n) = \mathcal{N} \exp\left[ - \frac{(n - n_{\rm path,end})^2}{\sigma_w^2} \right].
\end{equation}
Here, the effective width is typically set to $\sigma_w \approx 0.05\,(n_{\rm path,end} - n_{\rm path,  start})$. This specific profile ensures that deviations at intermediate densities are only weakly constrained—preserving the natural physical correlation of the GP prior. However, as the trajectory enters the final 5\% of the density span, the penalty increases exponentially that smoothly but forcefully merges the candidate path precisely into the pQCD constraints. Out of all generated candidates, we select the profile that yields the minimum action $\mathcal{S}_{\min}$. This least-action path naturally minimizes the boundary mismatch, serving as the basic skeleton for the GP conditioning.

The least-action path determines the general shape in $c_s^2(n)$, and it satisfies the required bounding integral \emph{approximately}. To guarantee \emph{exact} thermodynamic continuity, we process the skeleton through an exact GP condition projection in $\phi$-space.

We initialize the auxiliary profile using the skeleton associated with the least-action path, $\bm{\phi}^{(0)} = \bm{\phi}_{\rm skel}$. 
Since the continuous integral mapping from $c_s^2(n)$ to energy and pressure is nonlinear, we cannot apply standard one-step GP linear conditioning. Instead, we cast the exact matching as a non-linear Bayesian inverse problem \cite{tarantola2005inverse} and use a dynamically refined Newton-Raphson projection \cite{Rasmussen2006}.

At each iteration $k$, integrating the current density profile $\bm{\phi}^{(k)}$ yields an endpoint state $\bm{y}^{(k)}$ and a corresponding residual vector $\bm{r}^{(k)} = \bm{y}_{\rm tar} - \bm{y}^{(k)}$, where $\bm{y}$ represents the state vector of energy density and pressure. We linearize the constraint around the current path by computing the numerical Jacobian:

\begin{equation}
J^{(k)} = \frac{\partial \bm{y}}{\partial \bm{\phi}}\bigg|_{\bm{\phi}=\bm{\phi}^{(k)}}.
\label{eq:jacobian}
\end{equation}
To enforce this constraint while maximizing the probability relative to the GP prior, the required structural shift is computed using the bridge covariance matrix $K_{\rm br}$, following the iterative projection approach discussed above:

\begin{equation}
\delta\bm{\phi}^{(k)} = K_{\rm br}\,(J^{(k)})^\top\!\left(J^{(k)}\,K_{\rm br}\,(J^{(k)})^\top\right)^{-1}\!\bm{r}^{(k)}.
\label{eq:newton_proj}
\end{equation}
The profile is then updated via $\bm{\phi}^{(k+1)} = \bm{\phi}^{(k)} + \delta\bm{\phi}^{(k)}$. 

This iterative projection enforces the exact constraints through the GP covariance structure, preserving the physical smoothness dictated by $K_{\rm br}$. Convergence is achieved once the relative thermodynamic residuals satisfy the criteria $|\Delta\epsilon/\epsilon_{\rm tar,end}| < 10^{-3}$ and $|\Delta p/p_{\rm tar,end}| < 10^{-3}$. 

\subsection{How constraints shape the structure of $c_s^2$}
\label{sec:cs2_shape}

The inclusion of high-density pQCD limits imposes stringent restrictions on the EOS. While it is mathematically straightforward to construct arbitrary continuous paths in $P$–$n$ space that connect the pressure between {\it hadronic} and {\it pQCD} segment, these paths can lead to vastly different accumulated energy densities $\epsilon$.

\begin{figure}[htbp]
\centering
\vspace{-0.3cm}
\includegraphics[width=0.45\textwidth]{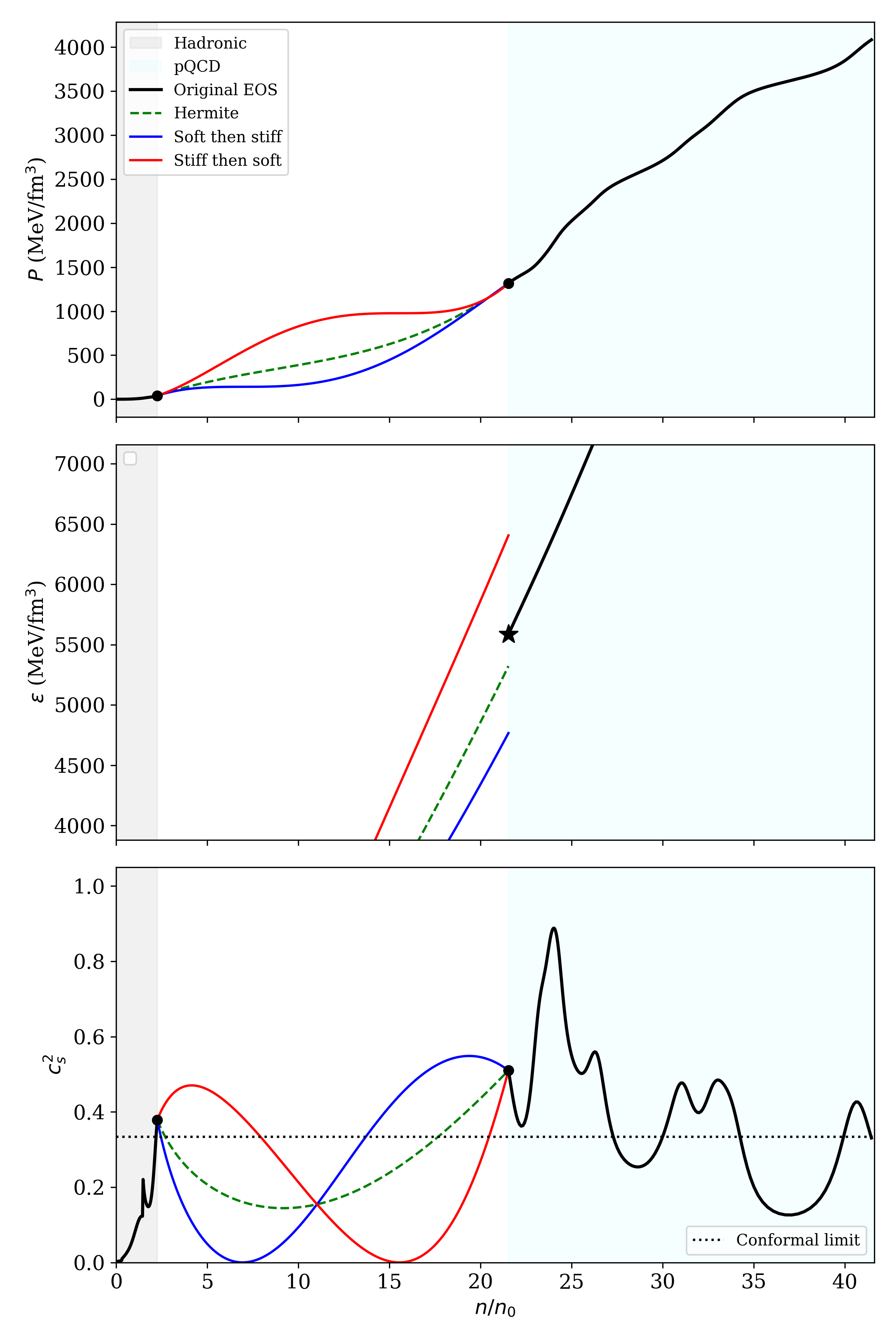}
\vspace{-0.3cm}
\caption{Schematic plot of the thermodynamic constraint on EOS. By tuning the amplitude $A$ in the shape function, three representative pressure (top) and corresponding $c_s^2$ (bottom) trajectories are constructed to exactly connect the pressure boundaries in {\it hadronic} and {\it pQCD} segment. While they share the same final pressure, the thermodynamic integral $\partial_n \epsilon = (\epsilon + P)/n$ dictates drastically different endpoint $\epsilon$ (middle).}
\label{fig:schematic_3path}
\end{figure}

To illustrate this systematically in Fig.~\ref{fig:schematic_3path}, we construct test pressure trajectories by using the same sample as in Fig.~\ref{fig:schematic} using a modified cubic Hermite spline (for simplicity, we maintain the continuity of $c_s^2$ here, but not that of its first derivative.):
\begin{equation}
    P(n) = P_{\rm Hermite}(n) + A \cdot B(x),
\end{equation}
where $x = (n-n_1)/(n_2-n_1)$ is the normalized density, $n_1$ and $n_2$ represent the boundary density for {\it hadronic} and {\it pQCD} segment. $P_{\rm Hermite}(n)$ is the standard cubic Hermite interpolation constrained by the boundary pressures and $c_s^2$. The shape function $B(x) = x^2(1-x)^2$ vanishes at the endpoints together with its first derivative. By tuning the amplitude $A$, we generate three representative paths that exactly match the $P$ boundaries:
\begin{itemize}
    \item \textbf{Hermite path ($A = 0$, green dashed):} A standard interpolation.
    \item \textbf{Soft-then-stiff path ($A < 0$, blue):} The EOS is initially softer than the standard interpolation, requiring a rapid stiffening near the high-density end.
    \item \textbf{Stiff-then-soft path ($A > 0$, red):} The EOS exhibits pronounced early stiffening, followed by a much softer state to meet the final pressure target.
\end{itemize}

As dictated by the thermodynamic relation $d\epsilon/dn = (\epsilon+P)/n$, these three paths yield drastically different energy-density accumulations even though they share the same final pressure.

To accommodate the existence of massive pulsars ($M \ge 2.0\,M_\odot$), astrophysical constraints demand that the EOS undergo prominent early stiffening, pushing its behavior toward the stiff-then-soft ($A > 0$) trajectory. However, this early stiffening also builds up $\epsilon$ very rapidly. If it persisted to higher densities, $\epsilon$ would systematically overshoot the tight pQCD bound.

Therefore, in order to simultaneously provide structural support for massive neutron stars and land precisely on the coupled $(P_{\rm pQCD},\epsilon_{\rm pQCD})$ point, any early stiffening in $c_s^2$ must be followed by a significantly softer state. This subsequent softening is a thermodynamic necessity that slows the accumulation of $\epsilon$. The non-monotonic structure of $c_s^2$ is not a mathematical artifact of a non-parametric method, but a fundamental global requirement that emerges when macroscopic astrophysical constraints are unified with microscopic pQCD limits.

\section{EOS and NS properties}
In this section, we provide extended details on the EOS properties and the corresponding macroscopic NS observables derived from our action-optimized Gaussian Process (GP) framework.

\begin{figure}[htbp]
\centering
\vspace{-0.3cm}
\includegraphics[width=0.45\textwidth]{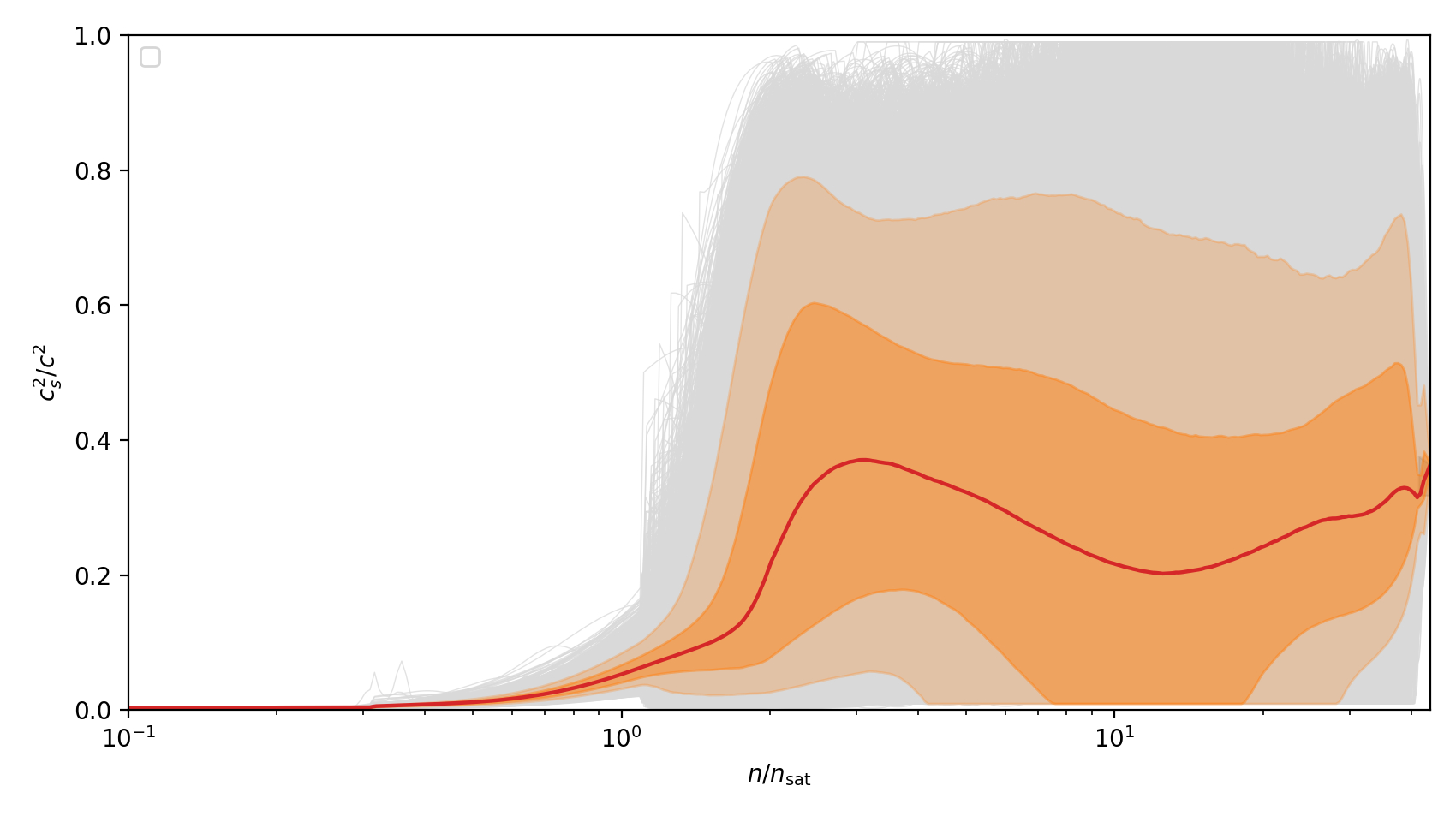}
\vspace{-0.3cm}
\caption{The prior EOSs continuously constructed from nuclear crust to asymptotic freedom.}
\label{fig:prior}
\end{figure}

We first present the prior distribution of $c_s^2(n)$, shown in Fig.~\ref{fig:prior}. Governed strictly by the auxiliary GP field $\phi(n)$, the prior spans a broad functional space that is globally shown towards a softer EOS at intermediate densities. Crucially, because the GP kernel enforces continuous, non-local correlations rather than rigid piecewise segments, our prior remains inherently smooth. It completely lacks the artificial, heavily localized structural biases or discontinuous derivatives that are often induced by traditional parametric methods (such as piecewise polytropes or fixed spectral representations). This smooth, unstructured baseline plays a vital role in our methodology: it ensures that the non-trivial structures—such as the prominent $c_s^2$ bumps and rapid stiffening—observed in the posterior are exclusively data-driven, emerging solely from the interplay between multi-messenger astrophysical constraints and strictly enforced asymptotic pQCD limits.

\begin{figure}[htbp]
\centering
\vspace{-0.3cm}
\includegraphics[width=0.45\textwidth]{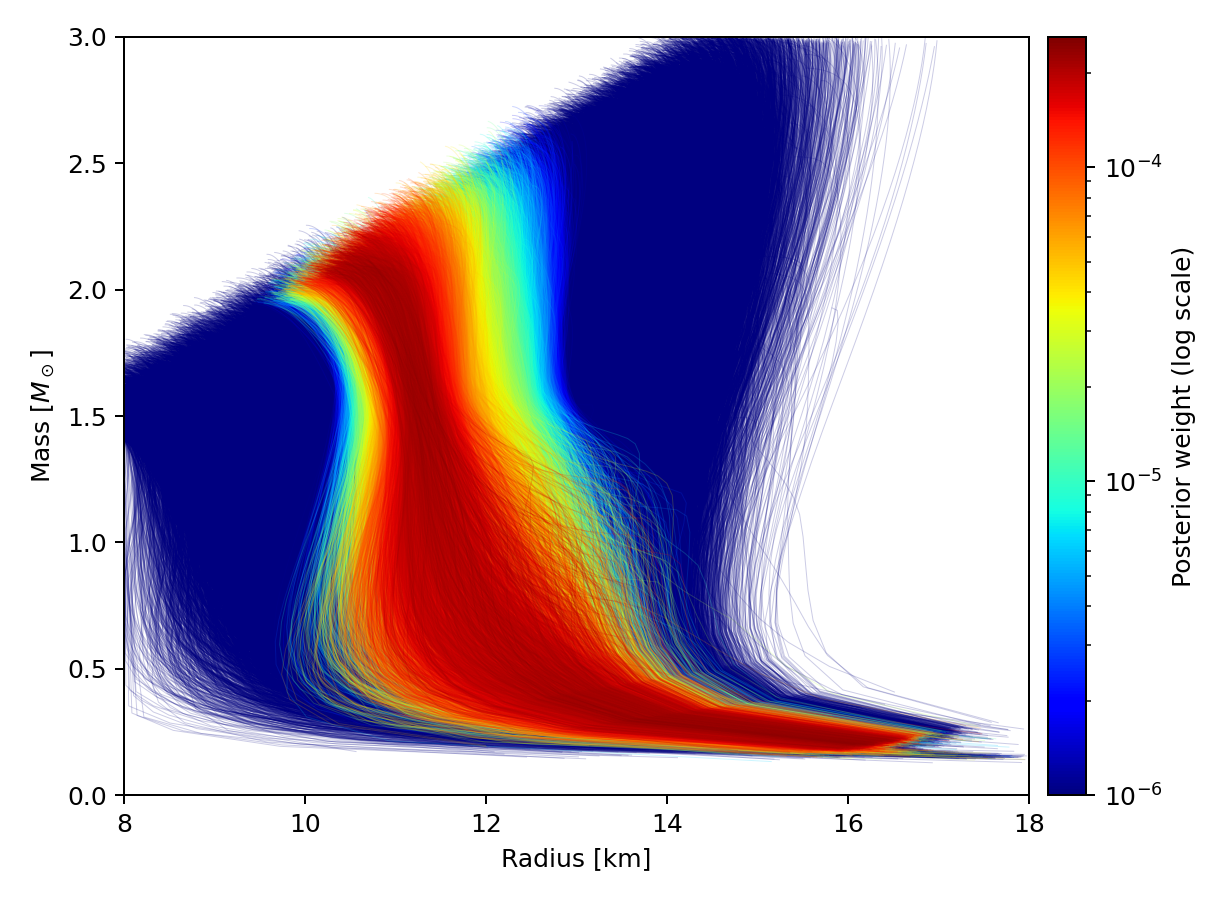}
\vspace{-0.3cm}
\caption{Posterior samples in the mass–radius plane, with color indicating the probability density.}
\label{fig:mr_constraint}
\end{figure}

In Fig.~\ref{fig:mr_constraint}, we present the posterior $M$--$R$ samples inferred from our conditioned EOS posterior ensemble. Each EOS trajectory is color-coded according to its statistical posterior weight, clearly highlighting the region of parameter space most favored by the data. These posterior draws are directly confronted with the primary multi-messenger constraints, including the NICER measurements for PSR~J0030+0451, PSR~J0740+6620, and PSR~J0437$-$4715, as well as the constraints from the GW170817. The dense concentration of highly weighted trajectories demonstrates that our action-optimized framework naturally provides the requisite intermediate-density stiffness to support heavy pulsars ($M \gtrsim 2.0\,M_\odot$), while simultaneously complying with the stringent compactness bounds imposed by lighter neutron stars.

\begin{figure}[htbp]
\centering
\vspace{-0.3cm}
\includegraphics[width=0.45\textwidth]{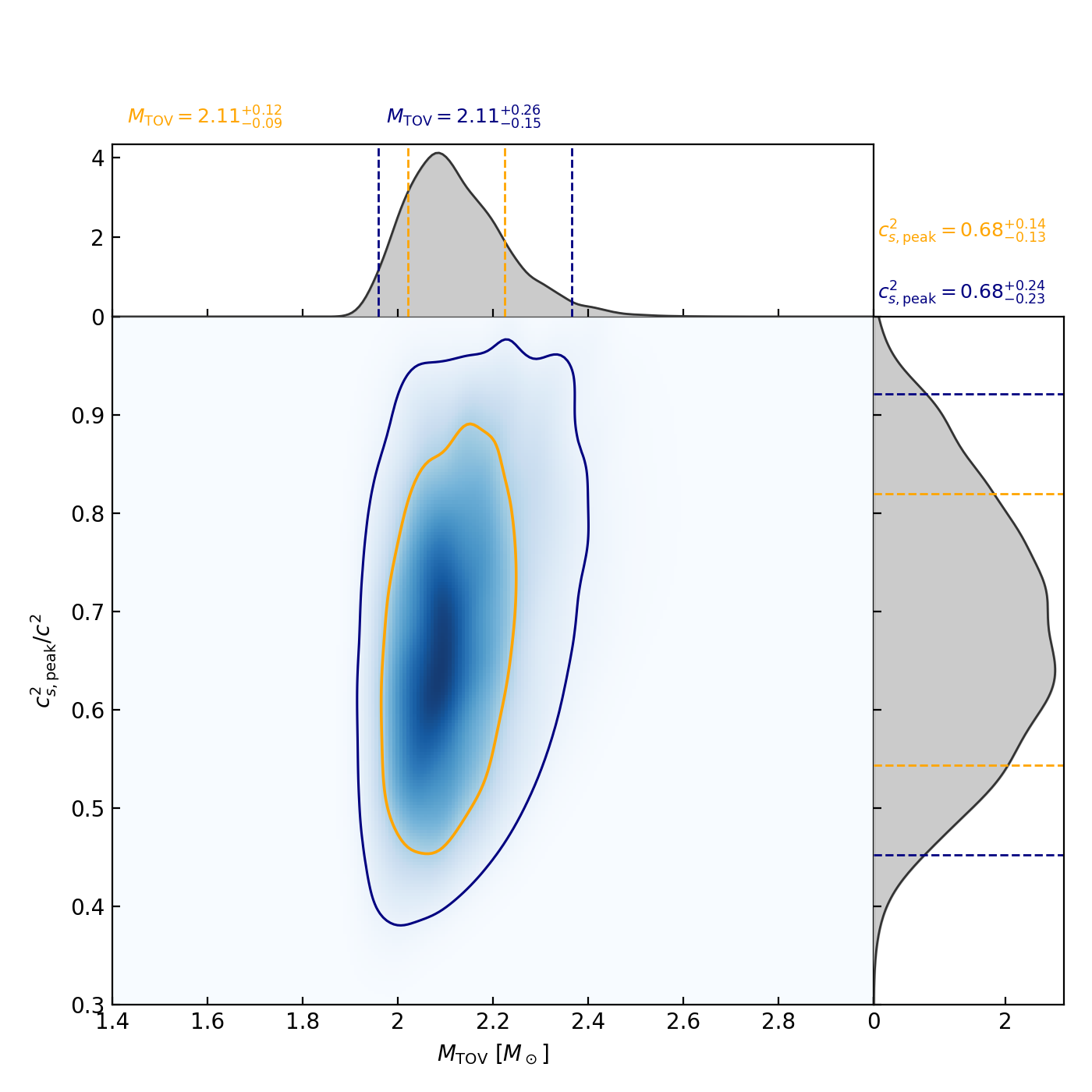}
\vspace{-0.3cm}
\caption{
Joint posterior distribution (68\% and 95\% contours) of $M_{\rm TOV}$ and the peak squared sound speed $c_{s,\rm peak}^2$. Their weak correlation demonstrates that a pronounced sound-speed peak is fully compatible with a moderate maximum mass, provided the EOS softens at higher densities.}
\label{fig:mtov_cs2_peak_joint}
\end{figure}

As discussed in the main text, the inferred sound-speed peak is only weakly correlated with $M_{\rm TOV}$, indicating that a pronounced intermediate-density stiffening does not require an exceptionally large maximum mass \cite{Shao:2019ioq,Han:2022rug,Fan:2023spm}. Fig.~\ref{fig:mtov_cs2_peak_joint} shows the full joint posterior distribution. The heaviest pulsar with a well-measured mass ($\sim 2.08\,M_\odot$ \cite{NANOGrav:2019jur}) is very close to the inferred $M_{\mathrm{TOV}} = 2.11^{+0.12}_{-0.09} \, M_{\odot}$ (with 95\% upper limit of $M_{\rm TOV}=2.37\,M_\odot$), implying that the dense matter cannot remain in a highly stiffened state over an extended density range. Instead, the sharp stiffening must be rapidly followed by softening, suggesting that the stiffening and softening phases originate from the same underlying physical process.

\begin{figure*}[htbp]
\centering
\vspace{-0.3cm}
\includegraphics[width=0.95\textwidth]{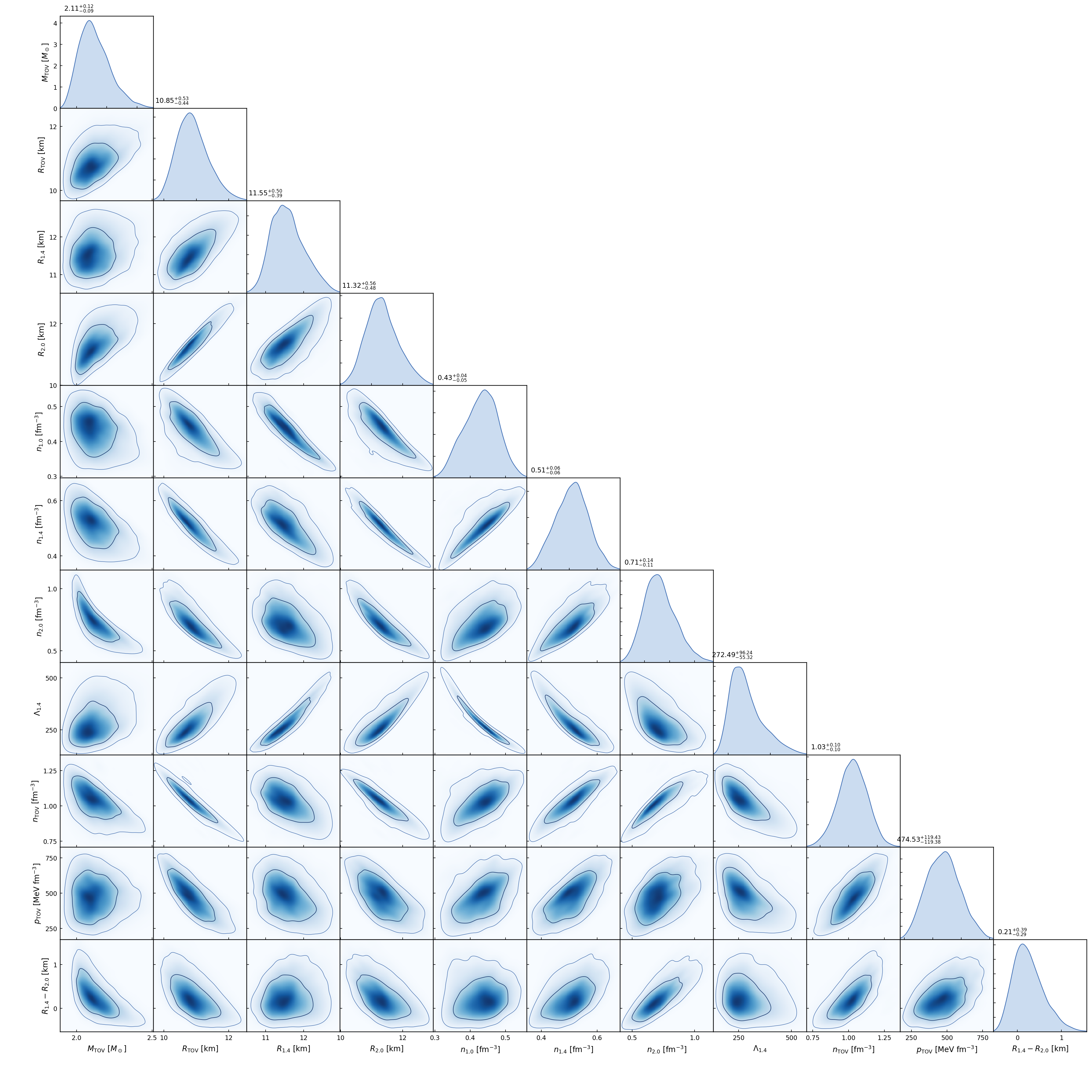}
\vspace{-0.3cm}
\caption{Posterior distribution for EOS parameters and NS properties. 68.3\% confidence interval is shown for each subplot.
}
\label{fig:corner}
\end{figure*}

Finally, Fig.~\ref{fig:corner} provides a comprehensive multi-variable corner plot exploring the joint posteriors among key macroscopic dimensions, including $M_{\mathrm{TOV}}$, some specific radii (e.g., $R_{1.4}$, $R_{2.0}$, $\Lambda_{1.4}$), and some EOS properties (e.g., $n_{\rm TOV}$, $P_{\rm TOV}$, $n_{1.4}$). This explicitly maps the covariances between the NS bulk properties, confirming that the internal EOS structures mapped by our method (e.g., the exact location and height of the sound speed bump) project consistently onto the correlated uncertainties of macroscopic NS observables. 

\section{Posterior with $M_{\rm TOV}$ reaching 2.5 $M_\odot$}
\begin{figure}[htbp]
\centering
\vspace{-0.3cm}
\includegraphics[width=0.45\textwidth]{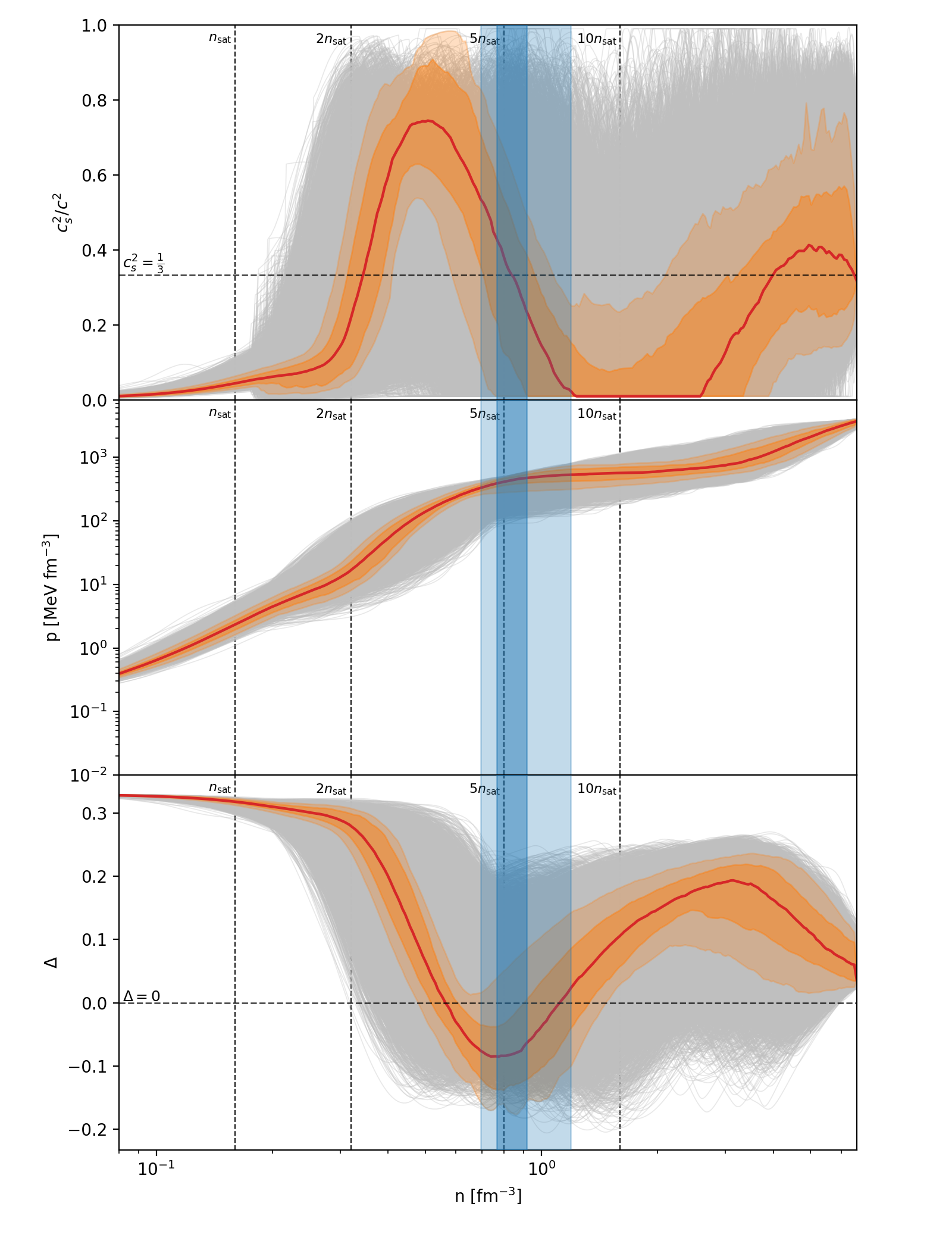}
\vspace{-0.3cm}
\caption{Similar as Fig.~\ref{fig:combo-n}, but here we assume that a $2.5 M_{\odot}$ NS was observed.}
\label{fig:m25}
\end{figure}

As discussed in the main text, accommodating a larger $M_{\mathrm TOV}$ demands stronger stiffening at intermediate densities. To push this thermodynamic compensation to its limit, we consider the case in which a $\sim 2.5 M_{\odot}$ NS is observed with $0.08 M_{\mathrm TOV}$ measurement uncertainty. With such an extreme mass constraint, the EOS must stiffen dramatically at lower densities to support the star against gravitational collapse. Consequently, in order to satisfy the pQCD condition without exceeding the energy-density limits, as Fig.~\ref{fig:m25} shows, the high-density EOS is thermodynamically driven to soften even more drastically at higher densities.

This behavior indeed contrasts with the conventional quark-star picture, where an intrinsically stiff quark EOS sustains massive compact objects. Our framework shows that, for a $\sim 2.5 M_{\odot}$ NS, the required quark-matter EOS becomes exceptionally soft instead. This counter-intuitive requirement thus offers a unique perspective for distinguishing neutron stars from quark stars.

\clearpage

\bibliographystyle{apsrev4-2}
\bibliography{refs}

\end{document}